\documentclass[preprint,preprintnumbers]{revtex4}
\usepackage{amssymb}
\usepackage{graphicx}
\usepackage{bm}
\begin{document}

\preprint{USTC-ICTS-12-03}

\title{Testing modified gravity models with recent cosmological observations}

\author{Wen-Shuai Zhang$^{1,2}$}
\email{wszhang@mail.ustc.edu.cn}

\author{Cheng Cheng$^{2,3}$}
\email{chcheng@itp.ac.cn}

\author{Qing-Guo Huang$^{2,3,4}$}
\email{huangqg@itp.ac.cn}

\author{Miao Li$^{2,3,4}$}
\email{mli@itp.ac.cn}

\author{Song Li$^{2,3}$}
\email{sli@itp.ac.cn}

\author{Xiao-Dong Li$^{1,2,5}$}
\email{renzhe@mail.ustc.edu.cn}

\author{Shuang Wang$^{1,2}$}
\email{swang@mail.ustc.edu.cn}

\affiliation{$^1$ Department of Modern Physics, University of Science and Technology of China, Hefei 230026, China}
\affiliation{$^2$ Institute of Theoretical Physics, Chinese Academy of Sciences, Beijing 100190, China}
\affiliation{$^3$ Kavli Institute for Theoretical Physics China, Chinese Academy of Sciences, Beijing 100190, China}
\affiliation{$^4$ State Key Laboratory of Frontiers in Theoretical Physics, Chinese Academy of Sciences, Beijing 100190, China}
\affiliation{$^5$ Interdisciplinary Center for Theoretical Study, University of Science and Technology of China, Hefei 230026, China}

\begin{abstract}
We explore the cosmological implications of five modified gravity (MG) models
by using the recent cosmological observational data,
including the recently released SNLS3 type Ia supernovae sample,
the cosmic microwave background anisotropy data from the Wilkinson Microwave Anisotropy Probe 7-yr observations,
the baryon acoustic oscillation results from the Sloan Digital Sky Survey data release 7,
and the latest Hubble constant measurement utilizing the Wide Field Camera 3 on the Hubble Space Telescope.
The MG models considered include the Dvali-Gabadadze-Porrati(DGP) model, two $f(R)$ models, and two $f(T)$ models.
We find that compared with the $\Lambda$CDM model,
MG models can not lead to a appreciable reduction of the $\chi^2_{min}$.
The analysis of AIC and BIC shows that the simplest cosmological constant model ($\Lambda$CDM) is
still most preferred by the current data, and the DGP model is strongly disfavored.
In addition, from the observational constraints, we also reconstruct the evolutions of the growth factor in these models.
We find that the current available growth factor data are not enough to
distinguish these MG models from the $\Lambda$CDM model.


\end{abstract}

\maketitle

\section{Introduction}\label{sec:intro}

Since its discovery in 1998 \cite{Riess},
the cosmic acceleration has become one of the central problems in theoretical physics and cosmology.
The most common approach to explain the cosmic acceleration is modifying the right hand side of Einstein equation,
i.e., introducing a mysterious composition called dark energy (DE) \cite{DEReview}.
Numerous theoretical DE models have been proposed in the last decade
\cite{quint}\cite{phantom}\cite{k}\cite{Chaplygin}\cite{tachyonic}\cite{refHDE}\cite{refADE}\cite{refRDE}\cite{hessence}\cite{YMC}\cite{Onemli},
while it is still as daunting as ever to understand the nature of the cosmic acceleration \cite{Weinberg}.

In addition to the introduction of the DE component,
another popular route to explain the cosmic acceleration,
i.e., modifying the left hand side of Einstein equation,
has also drawn a lot of attentions.
They are the so-called modified gravity (MG) models \cite{MGRNO,MGRFerr}.
So far, a lot of MG models have been proposed,
providing various interesting mechanisms for the cosmic acceleration.
For example, in the DGP braneworld model \cite{DGP},
gravity is altered at immense distances by slow leakage of gravity off from our three-dimensional universe,
leading to the apparent acceleration in the cosmological scale.
In the $f(R)$ gravity \cite{fR1,fR2},
the Ricci scalar $R$ is generalized to a function $f(R)$,
providing another possible mechanism for the cosmic acceleration.
A similar idea is the $f(T)$ gravity \cite{fT1,fT2,fTyang},
where the torsion $T$ is replaced by a function $f(T)$.
Other popular MG models include MOND \cite{MGMOND}, TeVeS \cite{MGTeVeS},
Brans-Dicke gravity \cite{MGBD}, Gauss-Bonnet gravity \cite{MGGB},
Horava-Lifshitz gravity \cite{MGHL}, and so on \cite{RTCahill}.
For recent reviews about MG models, see e.g. \cite{MGSF,MGRFerr}.

There are mainly two ways to constrain the cosmological models by using the observational data.
First, they can be tested through their effects in the cosmic expansion history,
which can be measured through Type Ia supernoave (SNIa),
baryon acoustic oscillations (BAO), and so on.
Second, they can be tested through their effects in the cosmic growth history,
which can be measured through the Integrated Sachs Wolfe (ISW) effect \cite{ISW},
weak lensing, galaxy clusters, and so on.
So far, many studies have been performed
\cite{DEManyModelsDRubin,Guo2006,MSa,MSb,MSc,MSd,Tsujikawa_Geff,Tsujikawa1_fR}.

In this work, we will mainly  test  the MG models against the observations of the cosmic expansion history.
Recently, a high-quality joint sample of 472 supernovae (SN), the SNLS3 SNIa dataset \cite{SNLS3Conley}, was released.
Moreover, the systematic uncertainties of the SNIa
data were nicely handled in work \cite{SNLS3Conley}.
It is interesting to study the performance of MG models when confronted with this newly released SNIa dataset.
Thus, in this paper we will adopt SNLS3 dataset and systematically study 5 representative MG models,
including DGP model, two $f(R)$ models, and two $f(T)$ models.
To perform a comprehensive analysis,
we also include the cosmic microwave background (CMB) anisotropy data from
the Wilkinson Microwave Anisotropy Probe 7-yr (WMAP7) observations \cite{WMAP7},
the baryon acoustic oscillation (BAO) results from the Sloan Digital Sky Survey
(SDSS) Data Release 7 (DR7) \cite{SDSSDR7},
and the latest Hubble constant measurement from the Wide Field Camera 3 (WFC3)
on the HST \cite{HSTWFC3}.
Utilizing these data, we constrain the parameter space of these MG models,
and compare them with the $\Lambda$CDM model by using the
AIC \cite{AIC} and BIC \cite{BIC} criterion.

An important tool to study the cosmic growth history is growth factor \cite{StarobinskyJETP1998}.
However, compared with the cosmic expansion history data,
the current growth factor observations have larger errors
(their redshifts are also not well determined).
Moreover, some data points of growth factor observations are obtained by assuming $\Lambda$CDM,
and therefore their use to test models different from $\Lambda$CDM may not be reliable.
So in this work we will not include the growth factor data
when performing observational constraints on the MG models.
Instead, making use of the constraints on the MG models
obtained from the observations of the cosmic expansion history,
we reconstruct the evolutions of the growth factor along with the redshift $z$ in these models
and compare the results with the growth factor data.
In this way, we can see whether it is possible to distinguish these MG models from the growth factor data.

This paper is organized as follows:
In Sec. II, we give a brief and overall description on the models considered here,
as well as the analysis methods used.
In Sec. III, we introduce in detail the current observational data of the cosmic expansion history
and the cosmic growth history, respectively.
In Sec. IV, we study the cosmological interpretations of the MG models considered here and give a summary of these models by making a comparison among them.
In Sec. V, we give a short discussion.
In addition, we present two appendixes about the initial condition of evolution equation of growth factor
and the effects of the parameter $n$ in $f(R)$ model.
In this work, we assume today's scale factor $a_{0}=1$, so the redshift $z=a^{-1}-1$;
the subscript ``0'' always indicates the present value of the corresponding quantity,
and the unit with $c=\hbar=1$ is used.

\section{Models and Methodology }

\subsection{Models}

In this work, we will investigate 5 representative MG models,
including

1. The DGP model

2. A power-law type $f(T)$ model ($f(T)_{PL}$)

3. An exponential type $f(T)$ model ($f(T)_{EXP}$)

4. The $f(R)$ model proposed by Starobinsky ($f(R)_{St}$)

5. The $f(R)$ model proposed by Hu and Sawicki ($f(R)_{HS}$)

The DGP model is a braneworld model \cite{DGP},
where gravity is altered at immense distances by slow leakage of gravity off from
our three-dimensional universe.
In the $f(T)$ models, the torsion $T$ in the Lagrangian density is replaced by a generalized function $T+f(T)$.
In this work, we consider two $f(T)$ models, in which the functions $f(T)$ are of
power-law type \cite{fT1} (hereafter $f(T)_{PL}$ model)
and exponential type \cite{fT2} (hereafter $f(T)_{EXP}$ model), respectively.
Similarly, in the $f(R)$ models, the Ricci scalar $R$ is replaced by $R+f(R)$,
and we will investigate the two popular $f(R)$ models proposed by Starobinsky \cite{fR2} (hereafter
$f(R)_{St}$ model) and by Hu and Sawicki \cite{fR1} (hereafter $f(R)_{HS}$ model), respectively.
We will give a detailed introduction about the formulas of these models in Sec. IV.

\subsection{The $\chi^2$ analysis}

To study the cosmological interpretations of the models listed in the previous subsection,
we have to make use of the data from the recent cosmological observations.
In practice, theoretical models and observational data can be related through the $\chi^2$ statistic.
For a physical quantity $\xi$ with experimentally measured value $\xi_{obs}$,
standard deviation $\sigma_{\xi}$ and theoretically predicted value $\xi_{th}$,
the $\chi^2$ takes the form
\begin{equation}
\chi^2_{\xi}={(\xi_{obs}-\xi_{th})^2\over \sigma^2_{\xi}}.
\end{equation}
In case that there are more than one independent physical quantities,
one can construct total $\chi^2$ by simply summing up all the $\chi^2_{\xi}$s, i.e.,
\begin{equation}
\chi^2=\sum_{\xi}\chi^2_{\xi}.
\end{equation}
In the following subsections,
we will present the detailed forms of the $\chi^2$ function of
the cosmological observations considered in this work.

Once having got the $\chi^2$ function,
one can determine not only the best-fit model parameters but also
the 1$\sigma$ and the 2$\sigma$ confidence level (CL) ranges
of the model considered by minimizing the total $\chi^2$ and
calculating $\Delta \chi^2 \equiv \chi^2-\chi^2_{min}$, respectively.
This need a thorough exploration of the parameter space of the $\chi^2$ function.
This procedure is commonly accomplished through the Monte Carlo Markov chain (MCMC) technique.
In this work, we make use of the publicly available CosmoMC package \cite{COSMOMC}
and generate $O(10^6)$ samples for each set of results presented in this paper.

\subsection{Model Comparison}

A statistical variable must be chosen to enforce a comparison between different models.
The $\chi _{min}^{2}$ may be the simplest one,
but it is difficult to compare models with different number of parameters.
Thus, in this work we use the commonly used AIC and BIC as model selection criterions.

The information criteria (IC) is the most frequently used method to assess different models.
It is also based on a likelihood method.
In this paper, we use the Akaike information criterion (AIC) \cite{AIC} and
the Bayesian information criterion (BIC) \cite{BIC} as model selection criteria.
These criteria favor models that have fewer parameters while giving a same fit,
and have been applied frequently in the cosmological contexts
\cite{Liddle:2004nh,cosmologyICGodlowski,cosmologyICBiesiada,cosmologyICMagueijo}.

The AIC \cite{AIC} is defined as
\begin{equation}
{\rm AIC}=-2\ln{\cal L}_{max} + 2 n_p,
\end{equation}
here ${\cal L}_{max}$ is the maximum likelihood and
$n_{p}$ denote the number of free model parameters.
Note that for Gaussian errors, ${\cal L}_{max}$ is related to the $\chi_{min}$ by
\begin{equation}
 -2\ln{\cal L}_{max}=\chi ^2_{min},
\end{equation}
so the difference in AIC can be simplified to $\Delta{\rm AIC}=\Delta\chi_{min}^2+ 2 \Delta n_p$.
As mentioned in Ref. \cite{Liddle:2007fy}, there is a version of the
AIC corrected for small sample sizes, ${\rm AIC}_c={\rm
AIC}+2n_p(n_p-1)/(N-n_p-1)$, which is important for $N/n_p\lesssim 40$.
In our case, this correction is negligible.

The BIC, also known as the Schwarz information criterion \cite{BIC}, is given by
\begin{equation}
{\rm BIC}=-2\ln{\cal L}_{max}+n_p\ln N,
\end{equation}
where $N$ is the number of data.
Also, in this case, the absolute value of the criterion is not of
interest, only the relative value between different models,
$\Delta{\rm BIC}=\Delta\chi_{min}^2+ \Delta n_p \ln N$, is useful.
it should be mentioned that the AIC is more lenient than BIC on models with
extra parameters for any likely data set $\ln N>2$.
Generally speaking,
a difference of 2 in BIC ($\Delta{\rm BIC}$) is considered as positive evidence
against the model with the higher BIC,
while a $\Delta{\rm BIC}$ of 6 is considered as strong evidence.

It should be noted that the IC alone can only imply that a more complex model is not necessary to explain the data,
since a poor information criterion result might arise from
the fact that the current data are too limited to constrain the extra parameters in this complex model,
and it might become preferred with the future data.
Thus, this only reflects the {\it current situation} for the theoretical models.

\section{Cosmological Observations}

In this section, firstly, we will introduce in detail
the observational data of the cosmic expansion history,
and then we will present the observational data of the cosmic growth history.

\subsection{Observations of the Cosmic Expansion History}

In this paper, we will constrain the MG models by using the observational data of the cosmic expansion history,
including the recently released SNLS3 SNIa sample \cite{SNLS3Conley},
the CMB anisotropy data from the WMAP7 observations \cite{WMAP7},
the BAO results from the SDSS DR7 \cite{SDSSDR7},
and the latest Hubble constant measurement utilizing the WFC3 on the HST \cite{HSTWFC3}.
In the following, we will   include these observations into the $\chi^2$ analysis.

\subsubsection{SNIa}

A most powerful probe of the cosmic acceleration is Type Ia supernovae (SNIa),
which are used as cosmological standard candles to directly measure the cosmic expansion.
Recently, a high-quality joint sample of 472 supernovae (SN),
the SNLS3 SNIa dataset \cite{SNLS3Conley}, was released.
This sample includes 242 SN at $0.08 < z < 1.06$ from
the Supernova Legacy Survey (SNLS) 3-yr observations \cite{Guy},
123 SN at low redshifts \cite{Hamuy}\cite{Hicken},
93 SN at intermediate redshifts from the Sloan Digital Sky Survey (SDSS)-II SN search \cite{Holtzman},
and 14 SN at $z > 0.8$ from Hubble Space Telescope (HST) \cite{HSTSNIa}.
This SNIa sample has been used to probe properties of DE  \cite{SNLS3Sullivan}\cite{SNLS3XDLi}\cite{SNLS3YWang},
to constrain the cosmological parameters \cite{SNLS3YGGong},
and to test the cosmological models \cite{SNLS3DEModels}.
Here, we briefly discuss the $\chi^2$ function of the SNLS3 SNIa data,
which can be downloaded from \cite{SNLS3Code}.

The $\chi^2$ function of the SNLS3 SNIa dataset takes the form
\begin{equation}
\chi^2_{\rm SN}=\Delta \overrightarrow{\bf m}^T \cdot {\bf C}^{-1} \cdot \Delta \overrightarrow{\bf m},
\end{equation}
where {\bf C} is a $472 \times 472$ covariance matrix
capturing the statistic and systematic uncertainties of the SNIa sample,
and $\Delta {\overrightarrow {\bf m}} = {\overrightarrow {\bf m}}_B - {\overrightarrow {\bf m}}_{\rm mod}$
is a vector of model residuals of the SNIa sample.
Here $m_B$ is the rest-frame peak $B$ band magnitude of the SNIa,
and $m_{\rm mod}$ is the predicted magnitude of the SNIa given by the cosmological model
and two other quantities (stretch and color) describing the light-curve of the particular SNIa.
The model magnitude $m_{\rm mod}$ is given by
\begin{equation}\label{SNchisq}
m_{\rm mod} = 5\log_{10} \mathcal{D}_L(z_{\rm hel},z_{\rm cmb})-\alpha(s-1)+\beta \mathcal{C} + \mathcal{M}.
\end{equation}
Here $\mathcal{D}_L$ is the Hubble-constant free luminosity distance.
In a spatially flat Friedmann-Robertson-Walker (FRW) universe
(the assumption of flatness is motivated by the inflation scenario), it takes the form
\begin{equation}
\mathcal{D}_L(z_{\rm hel}, z_{\rm zcmb}) = (1+z_{\rm hel})\int^{z_{\rm cmb}}_0 {{d z^\prime}\over{E(z^\prime)}}.
\end{equation}
Here $z_{\rm cmb}$ and $z_{\rm hel}$ are the CMB frame and heliocentric redshifts of the SN,
$s$ is the stretch measure for the SN,
and $\mathcal{C}$ is the color measure for the SN.
$E(z')\equiv H(z)/H_0$, and $H(z)$ is the Hubble parameter.
Notice that different model will give different $H(z)$ and $E(z)$.
$\alpha$ and $\beta$ are nuisance parameters which characterize the
stretch-luminosity and color-luminosity relationships, respectively.
Following \cite{SNLS3Conley}, we treat $\alpha$ and $\beta$ as free parameters and let them run freely.

The quantity $\mathcal{M}$ in Eq. (\ref{SNchisq}) is a nuisance parameter
representing some combination of the absolute magnitude of a fiducial SNIa and the Hubble constant.
In this work, we marginalize $\mathcal{M}$ following the complicated formula in Appendix C of \cite{SNLS3Conley}.
This procedure includes the host-galaxy information \cite{SullivanHostGalaxy} in the cosmological fits
by splitting the samples into two parts and allowing the absolute magnitude to be different between these two parts.

The total covariance matrix {\bf C} in Eq. (\ref{SNchisq})
captures both the statistical and systematic uncertainties of the SNIa data.
One can decompose it as \cite{SNLS3Conley},
\begin{equation}
{\bf C} = {\bf D}_{\rm stat} + {\bf C}_{\rm stat} + {\bf C}_{\rm sys},
\end{equation}
where ${\bf D}_{stat}$ is the purely diagonal part of the statistical uncertainties,
${\bf C}_{\rm stat}$ is the off-diagonal part of the statistical uncertainties,
and ${\bf C}_{\rm sys}$ is the part capturing the systematic uncertainties.
It should be mentioned that, for different $\alpha$ and $\beta$, these covariance matrices are also different.
Therefore, in practice one has to reconstruct the covariance matrix $\bf C$
for the corresponding values of $\alpha$ and $\beta$,
and calculate its inversion.
For simplicity, we do not describe these covariance matrices one by one.
One can refer to the original paper \cite{SNLS3Conley} and the public code \cite{SNLS3Code} for more details
about the explicit forms of the covariance matrices and the details of the calculation of $\chi^2_{\rm SN}$.

\subsubsection{CMB}

Then we turn to the CMB observational data.
We use the ``WMAP distance priors'' given by the 7-yr WMAP observations \cite{WMAP7}.
The distance priors include the ``acoustic scale'' $l_A$, the ``shift parameter'' $R$,
and the redshift of the decoupling epoch of photons $z_*$.
The acoustic scale $l_A$, which represents the CMB multipole
corresponding to the location of the acoustic peak,
is defined as \cite{WMAP7}
\begin{equation}
\label{ladefeq} l_A\equiv (1+z_*){\pi D_A(z_*)\over r_s(z_*)}.
\end{equation}
Here $D_A(z)$ is the proper angular diameter distance, given by
\begin{equation}
D_A(z)=\frac{1}{1+z}\int^z_0\frac{dz^\prime}{H(z^\prime)},
\label{eq:da}
\end{equation}
and $r_s(z)$ is the comoving sound horizon size, given by
\begin{equation}
r_s(z)=\frac{1} {\sqrt{3}}  \int_0^{1/(1+z)}  \frac{ da } { a^2H(a)
\sqrt{1+(3\Omega_{b0}/4\Omega_{\gamma0})a} },
\label{eq:rs}
\end{equation}
where $\Omega_{b0}$ and $\Omega_{\gamma0}$ are the present baryon
and photon density parameters, respectively.
In this paper,
we adopt the best-fit values, $\Omega_{b0}=0.02253 h^{-2}$
and $\Omega_{\gamma0}=2.469\times10^{-5}h^{-2}$ (for $T_{cmb}=2.725$ K),
given by the 7-yr WMAP observations \cite{WMAP7}.
Here $h$ is the reduced Hubble parameter satisfying $H_0 =  100h\ {\rm km/s/Mpc}$.
The fitting function of $z_*$ was proposed by Hu and Sugiyama \cite{Hu:1995en}:
\begin{equation}
\label{zstareq} z_*=1048[1+0.00124(\Omega_{b0}
h^2)^{-0.738}][1+g_1(\Omega_{m0} h^2)^{g_2}],
\end{equation}
where
\begin{equation}
g_1=\frac{0.0783(\Omega_{b0} h^2)^{-0.238}}{1+39.5(\Omega_{b0} h^2)^{0.763}},
\quad g_2=\frac{0.560}{1+21.1(\Omega_{b0} h^2)^{1.81}}.
\end{equation}
In addition, the shift parameter $R$ is defined as \cite{Bond97}
\begin{equation}
\label{shift} R(z_*)\equiv \sqrt{\Omega_{m0} H_0^2}(1+z_*)D_A(z_*),
\end{equation}
here $\Omega_{m0}$ is the present fractional matter density.

As shown in \cite{WMAP7}, the $\chi^2$ of the CMB data is
\begin{equation}
\chi_{CMB}^2=(x^{obs}_i-x^{th}_i)(C_{CMB}^{-1})_{ij}(x^{obs}_j-x^{th}_j),\label{chicmb}
\end{equation}
where $x_i=(l_A, R, z_*)$ is a vector,
and $(C_{CMB}^{-1})_{ij}$ is the inverse covariance matrix.
The 7-yr WMAP observations \cite{WMAP7} had given the maximum likelihood values:
$l_A(z_*)=302.09$, $R(z_*)=1.725$, and $z_*=1091.3$.
The inverse covariance matrix was also given in \cite{WMAP7}
\begin{equation}
(C_{CMB}^{-1})=\left(
  \begin{array}{ccc}
    2.305 & 29.698 & -1.333 \\
    29.698 & 6825.27 & -113.180 \\
    -1.333 & -113.180  &  3.414 \\
  \end{array}
\right).
\end{equation}

\subsubsection{BAO}

Next we consider the BAO observational data.
We use the distance measures from the SDSS DR7 \cite{SDSSDR7}.
One effective distance measure is the $D_V(z)$, which can be obtained
from the spherical average \cite{Eisenstein}
\begin{equation}
 D_V(z) \equiv \left[(1+z)^2D_A^2(z)\frac{z}{H(z)}\right]^{1/3},
\end{equation}
where $D_A(z)$ is the proper angular diameter distance.
In this work we use two quantities $d_{0.2}\equiv r_s(z_d)/D_V(0.2)$ and $d_{0.35}\equiv r_s(z_d)/D_V(0.35)$.
The expression of $r_s$ is given in Eq.(\ref{eq:rs}),
and $z_d$ denotes the redshift of the drag epoch, whose fitting formula
is proposed by Eisenstein and Hu \cite{BAODefzd}
\begin{equation}
\label{Defzd} z_d={1291(\Omega_{m0}h^2)^{0.251}\over 1+0.659(\Omega_{m0}h^2)^{0.828}}\left[1+b_1(\Omega_{b0}h^2)^{b2}\right],
\end{equation}
where
\begin{eqnarray}\label{Defb1b2}
b_1 &=& 0.313(\Omega_{m0}h^2)^{-0.419}\left[1+0.607(\Omega_{m0}h^2)^{0.674}\right], \\
b_2 &=& 0.238(\Omega_{m0}h^2)^{0.223}.
\end{eqnarray}
Following \cite{SDSSDR7}, we write the $\chi^2$ for the BAO data as,
\begin{equation}
\chi^2_{BAO}=\Delta{p_i}(C_{BAO}^{-1})_{ij}\Delta p_j,
\end{equation}
where
\begin{equation}
\Delta p_i = p^{\rm data}_i - p_i,
\ \  p^{\rm data}_1 = d^{\rm data}_{0.2 } = 0.1905,
\ \  p^{\rm data}_2 = d^{\rm data}_{0.35} = 0.1097,
\end{equation}
and the inverse covariance matrix takes the form
\begin{equation}
(C_{BAO}^{-1})=\left(
  \begin{array}{cc}
    30124  & -17227 \\
    -17227 & 86977 \\
  \end{array}
\right).
\end{equation}

\subsubsection{The Hubble Constant $H_0$}

We also use the prior on the Hubble constant $H_0$.
The precise measurements of $H_0$ will be helpful
to break the degeneracy between it and the DE parameters \cite{H0Freedman}.
When combined with the CMB measurement, it can lead to precise measurement of the DE EOS $w$ \cite{H0WHu}.
Recently, using the WFC3 on the HST,
Riess {\it et al.} obtained an accurate determination of the Hubble constant \cite{HSTWFC3}
\begin{equation}
H_0=73.8\pm 2.4 {\rm km/s/Mpc},
\end{equation}
corresponding to a $3.3\%$ uncertainty.
So the $\chi^2$ of the Hubble constant data is
\begin{equation}
\chi^2_{h}=\left({h-0.738\over 0.024}\right)^2.
\end{equation}

\subsubsection{Combining of the SNIa, CMB, BAO and $H_0$ data}

Since the SNIa, CMB, BAO and $H_0$ are effectively independent measurements,
we can combine them by simply adding together the $\chi^2$ functions,
i.e.,
\begin{equation}
\chi^2_{All} = \chi^2_{SN} + \chi^2_{CMB} + \chi^2_{BAO} + \chi^2_{h}.
\end{equation}

\subsection{Observations of the Cosmic Growth History}

As mentioned above, growth factor is an important tool to study the cosmic growth history.
In the following, we will give a brief introduction to the growth factor.

As shown in \cite{Boisseau_deltarho}, in the subhorizon limit,
the evolution equation of the matter density perturbation of MG theory has the following form,
\begin{equation}\label{eq:deltarho}
\ddot{\delta}+2H\dot{\delta}-4\pi G_{eff} \rho_{m}\delta=0,\hspace{10pt}
\end{equation}
where$\hspace{5pt} \delta\equiv\frac{\delta\rho_{m}}{\rho_{m} }$, $\rho_{m}$ is
the matter energy density and
$G_{eff}$ is the effective Newton's constant.
The explicit forms of $G_{eff}$ for the MG models considered in this paper
will be given in Sec.\ref{ModelsResults}.
The growth factor $f$ is defined as
\begin{equation}
f \equiv \frac{d\ln \delta}{d \ln a}.
\end{equation}
Thus, Eq.(\ref{eq:deltarho}) becomes
\begin{equation}\label{eq:deltagf}
\frac{df}{d\ln a}+f^2+(\frac{\dot{H}}{H}+2)f=\frac{3}{2} \frac{8\pi G_{eff}
\rho_{m}}{3H^2}.
\end{equation}
To solve Eq.(\ref{eq:deltagf}) numerically, the initial condition is taken
as $f(z=30)=1$ (see Appendix A).
Notice that different models correspond to different $H(z)$ and $G_{eff}$,
and thus give different $f(z)$.
We will make use of this equation to determine the predicted evolution history
of the growth factor in MG models.

To test the theoretical prediction of growth factor
obtained from Eq.(\ref{eq:deltagf}),
it may be helpful to consider the current growth factor data \cite{Porto2008,Nesseris2008,Guzzo2008},
which are summarized and listed in the Table \ref{gfdata}.
Compared with the cosmic expansion history data,
the current growth factor observations have larger errors,
and their redshifts are also not well determined.
Moreover, as pointed out in \cite{Nesseris2008},
some data points of Table \ref{gfdata} are obtained by assuming $\Lambda$CDM
when converting redshifts to distances for the power spectra,
and therefore it may be unreliable to use them in the models different from $\Lambda$CDM.
So in this work, we will not include the growth factor data in our $\chi^2$ analysis.
Instead, by using the constraints obtained from the cosmic expansion history data,
we will explore the evolutions of these models' growth factor along with redshift $z$,
and compare the results with the growth factor data.

\begin{table}[htp]
\begin{center}
\begin{tabular}{|lcr|}
\hline
\ \ \ \ \ \ \ \ \   $z$\ \ \ \ \ \ \ \ \   & \ \ \ \ \ \ \  $f_{obs}$\ \ \ \ \ \ \  &\ \ References \\
\hline
$0.15$ & $0.49\pm 0.1$ & \cite{Colless2001,Guzzo2008} \\
$0.35$ & $0.7\pm 0.18$ & \cite{Tegmark2006} \\
$0.55$ & $0.75\pm 0.18$ & \cite{Ross2007} \\
$0.77$ & $0.91\pm 0.36$ & \cite{Guzzo2008} \\
$1.4$ & $0.9\pm 0.24$ & \cite{ngela2008} \\
$3.0$ & $1.46\pm 0.29$ & \cite{McDonald2005} \\
$2.125-2.72$ & $0.74\pm 0.24$ & \cite{Viel2004} \\
$2.2-3$ & $0.99\pm 1.16$ & \cite{Viel2006}\\
$2.4-3.2$ & $1.13\pm 1.07$ & \cite{Viel2006}\\
$2.6-3.4$ & $1.66\pm 1.35$ & \cite{Viel2006} \\
$2.8-3.6$ & $1.43\pm 1.34$ & \cite{Viel2006} \\
$3-3.8$ & $1.3\pm 1.5$ & \cite{Viel2006}\\
\hline
\end{tabular}
\end{center}
\caption{The combined observational data on the growth factor $f$.}
\label{gfdata}
\end{table}

\section{Results and Discussions} \label{ModelsResults}

In the following, we firstly constrain the parameter spaces of the models
by using the observational data of the cosmic expansion history,
then we plot the corresponding evolutions of these models' growth factor along with redshift $z$,
and compare the results with the observational data of growth factor $f_{obs}$.

\subsection{DGP model}

We start with the flat DGP model \cite{DGP}.
In this model, the cosmic expansion history $E(z)\equiv H(z)/H_0$ is determined by the following equation \cite{DGP_Hz}:
\begin{eqnarray}
E(z) = \left[ \sqrt{\Omega_{m0}(1+z)^3+\Omega_{r0}(1+z)^4+\Omega_{rc}}+\sqrt{\Omega_{rc}}\right] .
\end{eqnarray}
Here, $\Omega_{rc} = (1-\Omega_{m0}-\Omega_{r0})^2/4$,
and $\Omega_{r0}$ is the present fractional radiation density given by \cite{WMAP7}
\begin{equation}
\Omega_{r0}=\Omega_{\gamma0}(1+0.2271N_{eff}),\ \ \
\Omega_{\gamma0}=2.469\times10^{-5}h^{-2},\ \ \ N_{eff}=3.04,
\end{equation}
where $\Omega_{\gamma0}$ is the present fractional photon density,
and $N_{eff}$ is the effective number of neutrino species.

In this model, the effective Newton constant $G_{eff}$ is no longer the simple constant $G$,
instead, it takes the following form \cite{DGP_Geff},
\begin{equation}\label{eq:DGP_Geff}
G_{eff}=G\frac{2(1+2\Omega_{m}^2(z))}{3(1+\Omega_{m}^2(z))}.
\end{equation}

\begin{figure}
\includegraphics[scale=0.59, angle=0]{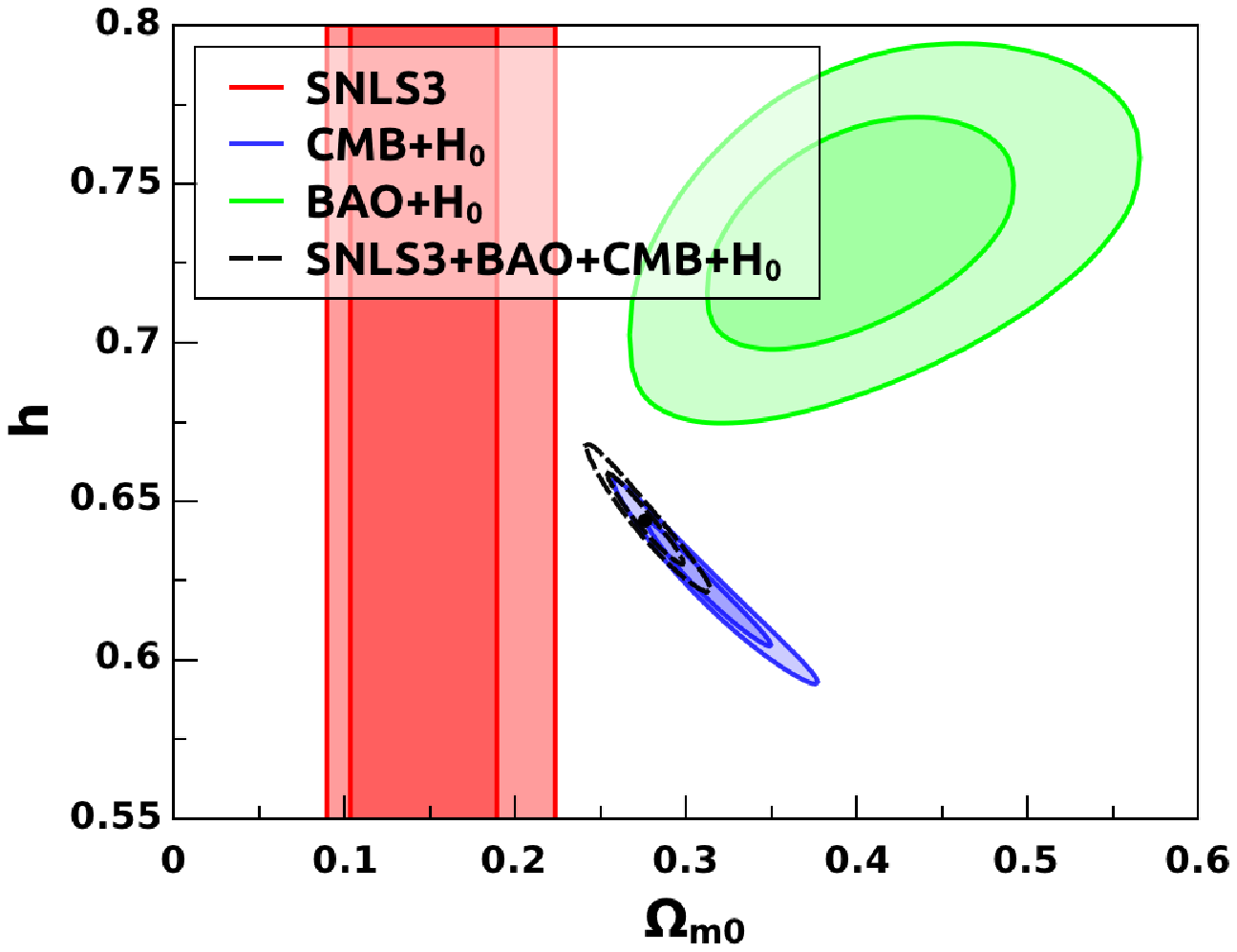}
\includegraphics[scale=0.58, angle=0]{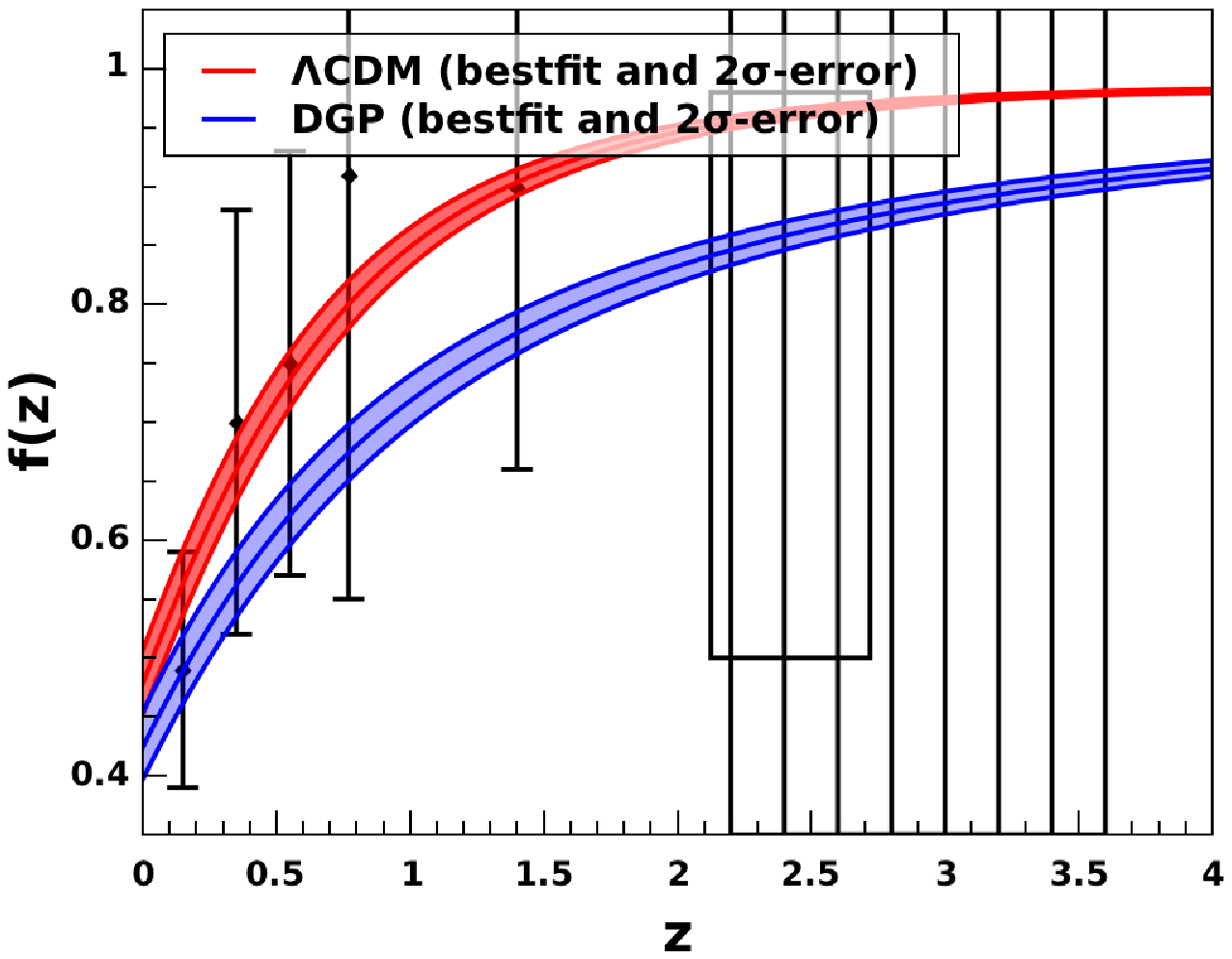}
\caption{\label{DGPfg}
$Left Panel$: Probability contours at the $68.3\%$ and $95.4\%$ confidence levels
in $\Omega_{m0} \textendash h$ plane, for the DGP model.
This figure shows that there is an inconsistency between different cosmological probes for DGP model.
$Right\ Panel$: The evolutions of $f(z)$ predicted by the $\Lambda$CDM and DGP models.
The best-fit values and $2\sigma$ errors regions determined by SNIa+CMB+BAO+$H_{0}$ analysis are shown.}
\end{figure}

Like the $\Lambda$CDM model, DGP is also an one-parameter model with the sole parameter $\Omega_{m0}$.
However, it has been shown to be disfavored by the cosmological observations \cite{DGP_disfavor,Guo2006,DEManyModelsDRubin}.
In this work we obtain similar result.
In the left panel of Fig.\ref{DGPfg}, we plot the contours of $68.3\%$ and $95.4\%$ confidence levels
in the $\Omega_{m0} \textendash h$ plane, for the DGP model.
Constraints from SNLS3, CMB+$H_0$, BAO+$H_0$, and SNLS3+CMB+BAO+$H_0$ are shown in contours with different colors.
The figure shows that there is an inconsistency between different cosmological probes in the DGP model.
This also implies that the DGP model is disfavored by these cosmological probes.
At the $68.3\%$ and $95.4\%$ confidence levels, we get
\begin{equation}
\Omega_{m0} = 0.2753^{+0.0151}_{-0.0140}\ ^{+0.0312}_{-0.0274},
\ \ \ h = 0.6445^{+0.0093}_{-0.0093}\ ^{+0.0189}_{-0.0186}.
\end{equation}

In the right panel of Fig.\ref{DGPfg}, we plot the evolutions of $f(z)$
predicted by the $\Lambda$CDM model and the DGP model
along with the observational data of growth factor $f_{obs}$.
From this figure, we can see clear difference between
the predicted evolutions of $f(z)$ from the DGP model and the $\Lambda$CDM model.
However, the current available growth factor data are still
not powerful enough to distinguish these two models.

\subsection{$f(T)$ models}

Then, we turn to the $f(T)$ models. The action of $f(T)$ models is
\begin{equation}
S = \frac{1}{16 \pi G} \int d^{4}x\sqrt{-g}\left[ T+f(T) \right]  + S_{m} + S_{r},
\end{equation}
here, $S_{m}$ and $S_{r}$ are the action of the matter content and the radiation content, and $T$ is the torsion scalar \cite{fT1}.

Assuming a flat homogeneous and isotropic FRW universe, $T$ satisfies
\begin{equation}
T=6H^2,
\end{equation}
and the modified Friedmann equations are obtained as \cite{fT2}
\begin{eqnarray}
\label{eq:Friedmann1fT}
H^2&=&\frac{8\pi G}{3}(\rho_m + \rho_r )-\frac{f}{6}-2H^2f_{T}, \\
\dot{H}&=&-\frac{1}{4}\frac{6H^2+f+12H^2f_{TT}}{1+f_{T}-12H^2f_{TT}},
\end{eqnarray}
here and throughout, $\rho_{m}$ and $\rho_{r}$ are matter density and radiation density.
$f_{T}$ and $f_{TT}$ are defined as
\begin{equation}
f_{T}\equiv \frac{df}{dT}, \hspace{10pt} f_{TT}\equiv \frac{d^2f}{dT^2}.
\end{equation}

For $f(T)$ models, $G_{eff}$ is given by \cite{fT_Geff}
\begin{equation} \label{eq:fTGeff}
 G_{eff}=\frac{G}{1+f_{T}} .
\end{equation}
Notice that, if $f(T)$ is a constant, the term $f(T)$ acts just like a cosmological constant.

In this paper,
we consider two $f(T)$ models with different types of parametrization:
One is the power law model considered in \cite{fT1},
the other is exponential form model proposed by Linder \cite{fT2}.
For simplicity, hereafter we will call them $ f(T)_{PL}$ and $ f(T)_{EXP}$, respectively.

\subsubsection{The $f(T)_{PL}$ model}

The power law model \cite{fT1} assumes the following ansatz of $f(T)$,
\begin{equation} \label{eq:fT1}
f(T) = \alpha (-T)^{n}.
\end{equation}
Here $\alpha$ can be obtained by matching the present matter density $\Omega_{m0}$ \cite{fT1} :
\begin{equation}
\alpha = (6 H_{0}^{2})^{1-n} \frac{1-\Omega_{m0}}{2n-1}.
\end{equation}
Making use of these two equations, Eq.(\ref{eq:Friedmann1fT}) can be written as
\begin{equation}
E(z)^2 = \Omega_{m0}(1+z)^3+\Omega_{r0}(1+z)^4+(1-\Omega_{m0}-\Omega_{r0}) E(z)^{2n}.
\end{equation}
Correspondingly, we also obtain  $G_{eff}$ :
\begin{equation}
G_{eff} = G\frac{1}{1+\frac{n(1-\Omega_{m0})}{1-2n}(\frac{H_{0}}{H})^{2(1-n) } }.
\end{equation}

\begin{figure}
\includegraphics[scale=0.58, angle=0]{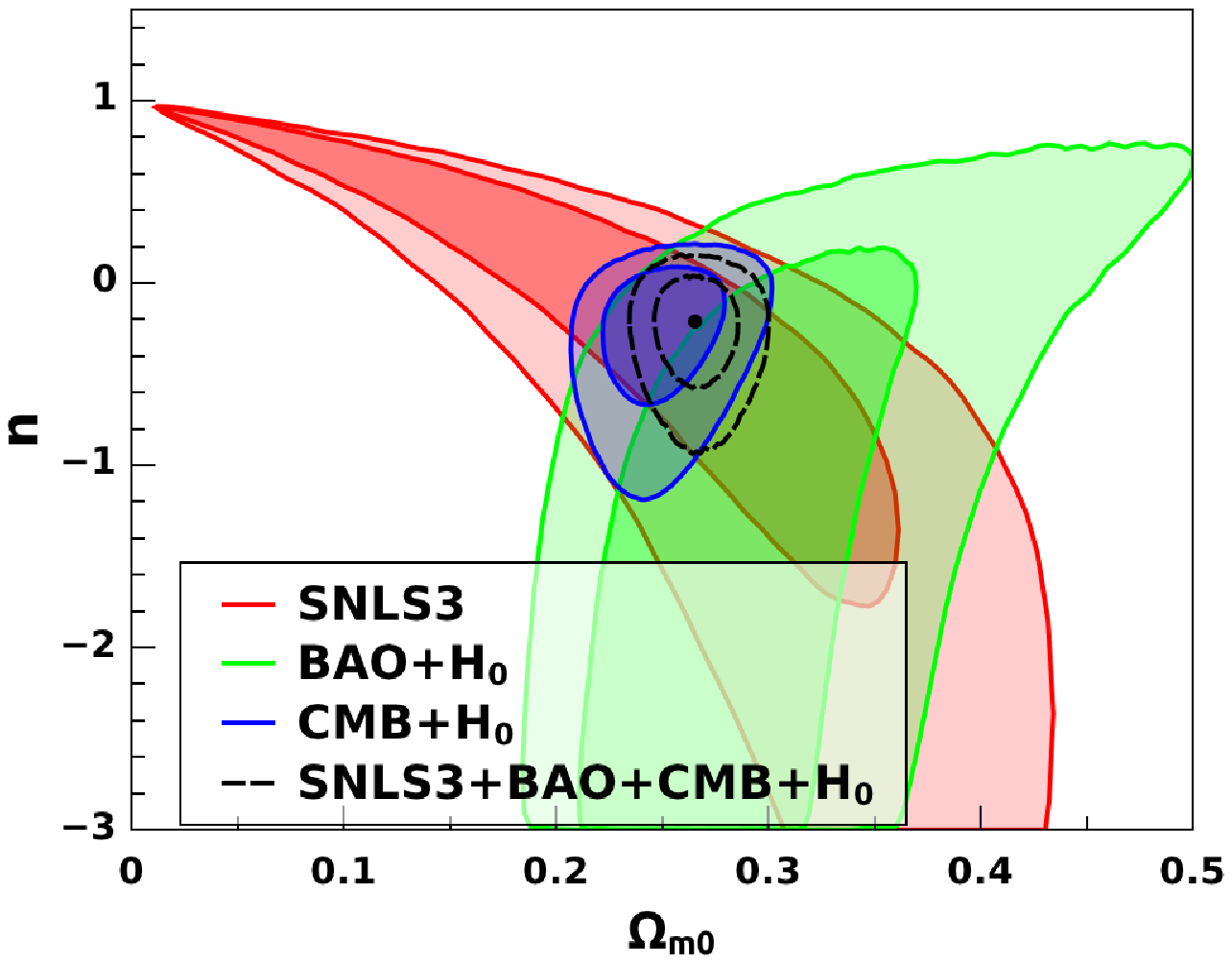}
\includegraphics[scale=0.59, angle=0]{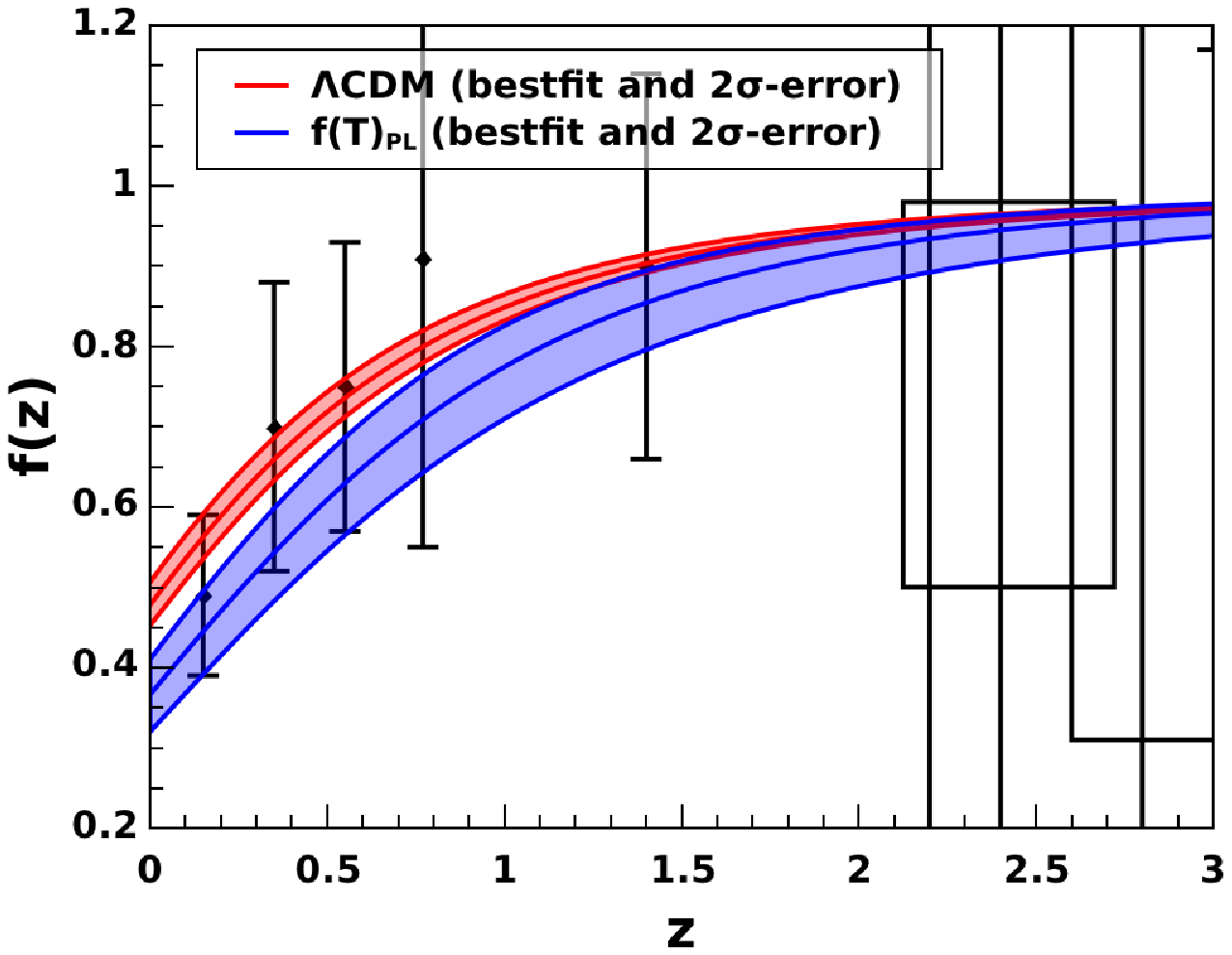}
\caption{\label{fT1fg}
$Left\ Panel$: Probability contours at the $68.3\%$ and $95.4\%$ confidence levels
in $\Omega_{m0} \textendash n$ plane, for the $f(T)_{PL}$ model.
$Right\ Panel$: The evolutions of $f(z)$ predicted by the $\Lambda$CDM and $f(T)_{PL}$ models.
The best-fit values and $2\sigma$ errors regions determined by SNIa+CMB+BAO+$H_{0}$ analysis are shown.
}
\end{figure}

Unlike the $\Lambda$CDM model or the DGP model,
this model has two model parameters $n$ and $\Omega_{m0}$.
In the left panel of Fig.\ref{fT1fg}, we plot the contours of $68.3\%$ and $95.4\%$ confidence levels
in the $\Omega_{m0} \textendash n$ plane, for the $f(T)_{PL}$ model.
Constraints from SNLS3, CMB+$H_0$, BAO+$H_0$, and SNLS3+CMB+BAO+$H_0$ are shown in contours with different colors.
At the $68.3\%$ and $95.4\%$ confidence levels,
\begin{equation}
\Omega_{m0} = 0.2652^{+0.0208}_{-0.0190}\ ^{+0.0356}_{-0.0308},
\ \ \ n = -0.2057^{+0.2490}_{-0.3714}\ ^{+0.3663}_{-0.7433},
\ \ \ h = 0.7243^{+0.0228}_{-0.0222}\ ^{+0.0376}_{-0.0366}.
\end{equation}
Unlike the DGP model, the different contours given by different observations overlap,
showing a consistent fit.

In the right panel of Fig.\ref{fT1fg}, we plot the evolutions of $f(z)$
predicted by the $\Lambda$CDM model and the $f(T)_{PL}$ model
along with the observational data of growth factor $f_{obs}$.
We see that the predicted evolutions of $f(z)$ in these two models slightly differ from each other in the 2$\sigma$ CL,
especially at the low-redshift region. But the current growth factor data
are still not powerful enough to distinguish these two models.

\subsubsection{The $f(T)_{EXP}$ model}

Another popular $f(T)$ model is of exponential form model proposed by Linder \cite{fT2}.
It takes the form
\begin{equation}
f(T) = c T_{0}(1-\exp(-p\sqrt{T/T_{0}})),
\end{equation}
with
\begin{equation}
c = \frac{1-\Omega_{m0}}{1-(1+p)^{-p}},\hspace{10pt} T_{0} = 6H_{0}^{2}.
\end{equation}
Combining the above two equations with Eqs.(\ref{eq:Friedmann1fT}-\ref{eq:fTGeff}),
after some tedious calculation,
one can obtain the $E(z)$ and $G_{eff}$ of this model,
\begin{equation}
E^2(z) = \Omega_{m0}(1+z)^3+\Omega_{r0}(1+z)^4+c[1-(1+pE(z))\exp(-pE(z))],
\end{equation}
\begin{equation}
G_{eff} =  \frac{1}{1-\frac{p (p+1)^p e^{-\sqrt{E(z)} p}
\left(\Omega_{m0}-1\right)}{2 \sqrt{E(z)} \left((p+1)^p-1\right)}}.
\end{equation}

\begin{figure}
\includegraphics[scale=0.58, angle=0]{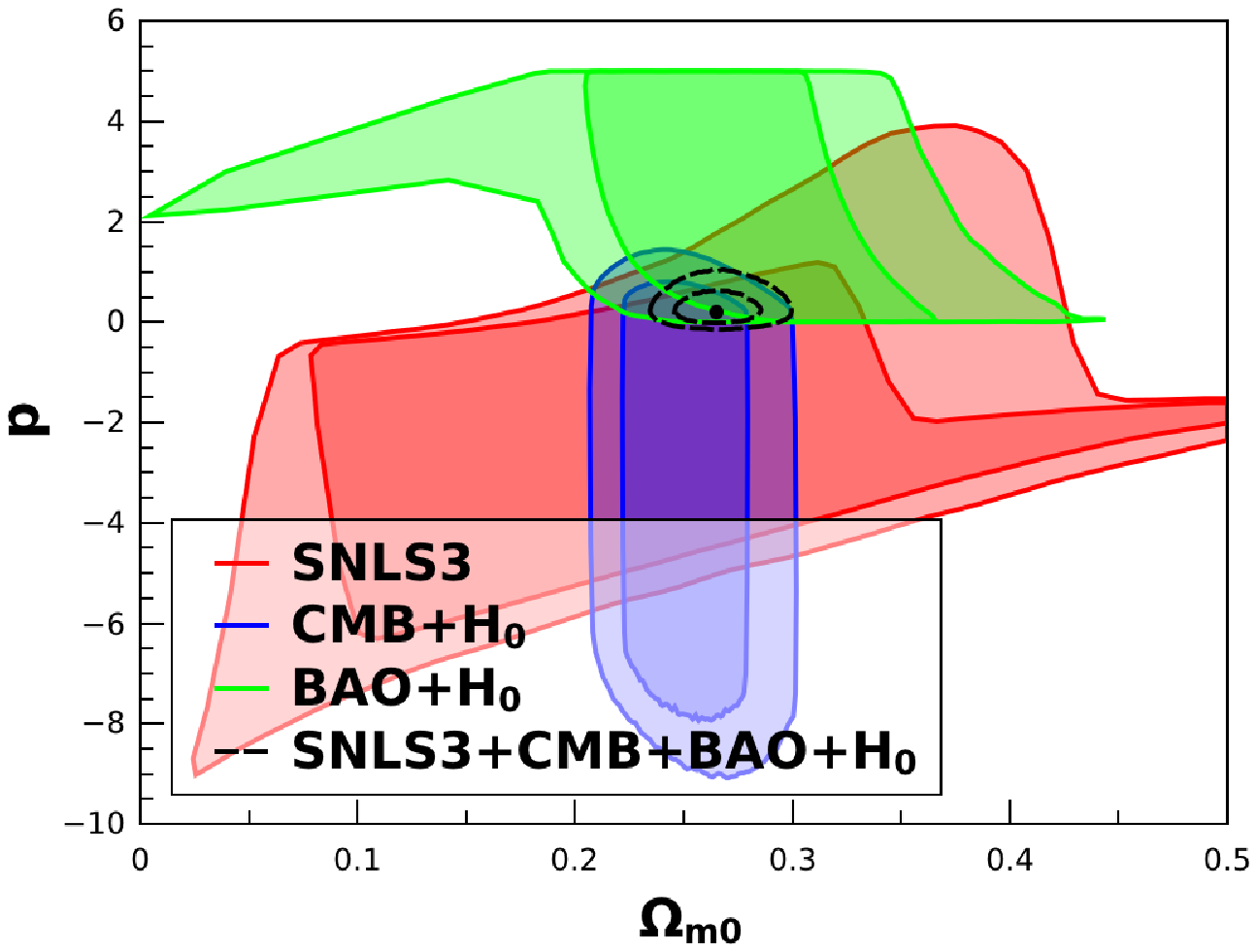}
\includegraphics[scale=0.58, angle=0]{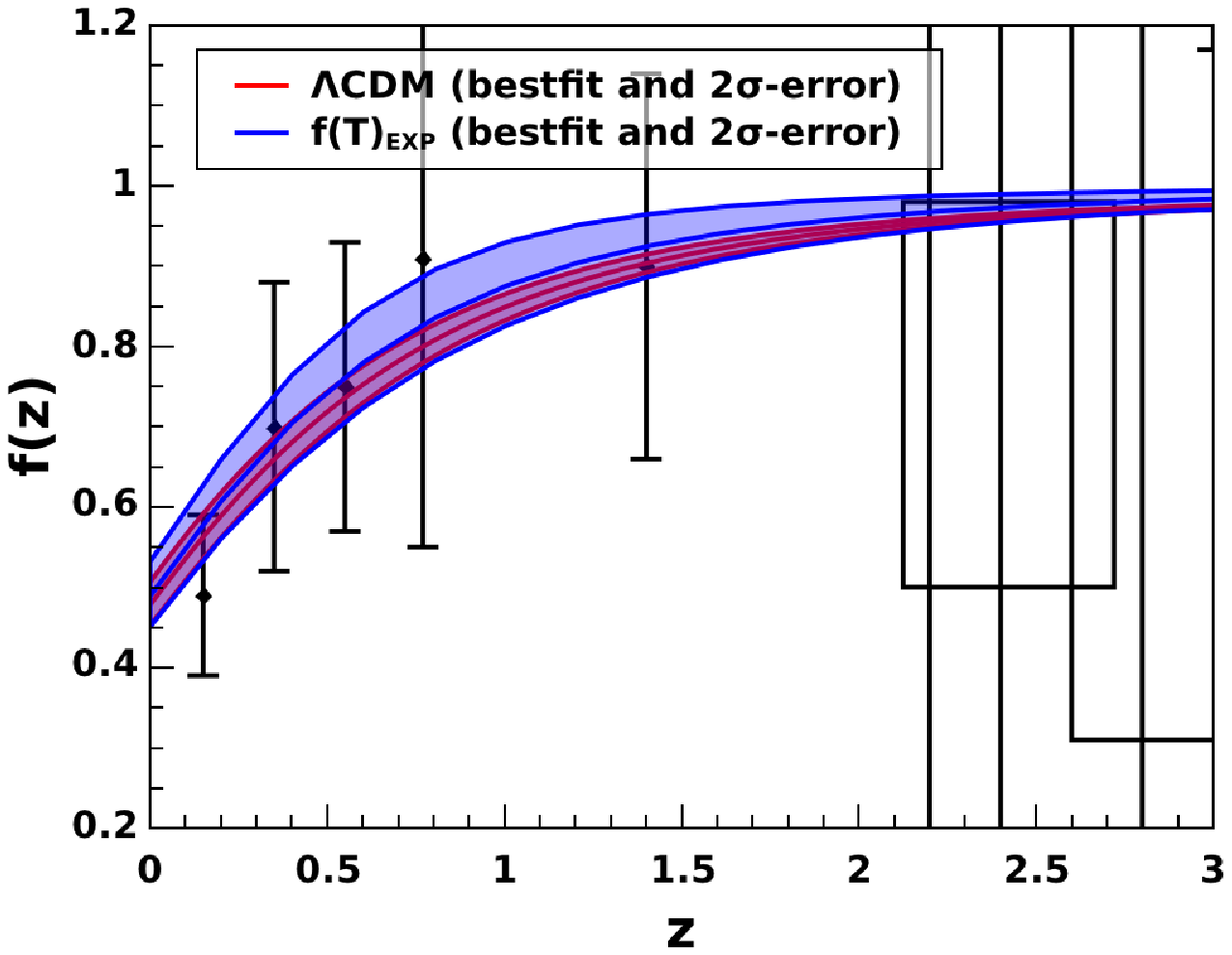}
\caption{\label{fT2fg}
$Left\ Panel$: Probability contours at the $68.3\%$ and $95.4\%$ confidence levels
in $\Omega_{m0} \textendash p$ plane, for the $f(T)_{EXP}$ model.
$Right\ Panel$: The evolutions of $f(z)$ predicted by the $\Lambda$CDM and $f(T)_{EXP}$ models.
The best-fit values and $2\sigma$ errors regions determined by SNIa+CMB+BAO+$H_{0}$ analysis are shown.}
\end{figure}

Like the $f(T)_{PL}$ model,
this model is also a two-parameter model with parameters $p$ and $\Omega_{m0}$.
The contours of the $f(T)_{EXP}$ model are shown in the left panel of Fig.\ref{fT2fg}.
As before, constraints from different cosmological data are shown in contours with different colors.
At the $68.3\%$ and $95.4\%$ confidence levels, we obtain
\begin{equation}
\Omega_{m0} = 0.2646^{+0.0210}_{-0.0192}\ ^{+0.0355}_{-0.0305},
\ \ \ p = 0.2301^{+0.4044}_{-0.2537}\ ^{+0.8346}_{-0.3788},
\ \ \ h = 0.7241^{+0.0223}_{-0.0216}\ ^{+0.0366}_{-0.0354}.
\end{equation}

In the right panel of Fig.\ref{fT2fg}, we plot the evolutions of $f(z)$
predicted by the $\Lambda$CDM model and the $f(T)_{EXP}$ model
along with the observed data of growth factor $f_{obs}$.
This figure shows that the predicted evolutions $f(z)$ from the expansion history data overlap.
Especially, the constrained region of $f(z)$ in the $\Lambda$CDM model is a subset of that of the $f(T)_{EXP}$ model.
This means that it will be more difficult to distinguish the cosmic growth history of these two models from the growth factor data.

\subsection{$f(R)$ models}

At last, let us investigate the $f(R)$ model. The basic idea of this model is replacing $R$ by $R + f(R)$,
yielding the action
\begin{equation}
S = \frac{1}{16 \pi G} \int d^{4}x\sqrt{-g} \left[ R+f(R) \right] + S_{m} + S_{r},
\end{equation}
where $S_{m}$ and $S_{r}$ are the actions for the matter content and the radiation content respectively.
For the background FRW metric, the Ricci scalar can be determined by the Hubble parameter and its time derivative, i.e.,
\begin{equation}
R = 12 H^{2} + 6 \dot{H}.
\end{equation}
Taking variations of the action of $f(R)$ model
with respect to the metric in the spatially flat FRW universe,
one can obtain the modified Friedmann equation:
\begin{eqnarray}
\label{eq:Friedmann1fR}
H^{2} - f_{R}(\frac{R}{6}-H^{2}) + \frac{f}{6} + H^{2}f_{RR}R^{'}
= \frac{8 \pi G}{3}(\rho_m + \rho_r ).
\end{eqnarray}
Here, primes denote derivatives with respect to $\ln a$,
and $f_{R}$ and $f_{RR}$ are defined by
\begin{equation}
f_{R} \equiv \frac{df}{dR}, \ \ \  f_{RR} \equiv \frac{d^{2}f}{dR^{2}}.
\end{equation}
The $G_{eff}$ of the $f(R)$ model is given by \cite{Tsujikawa_Geff}
\begin{equation}
\label{eq:fRGeff}
G_{eff}=\frac{G}{1+f_{R}}\frac{1+4\frac{k^2}{a^2}\frac{f_{RR}}{1+f_{R}}}
{1+3\frac{k^2}{a^2}\frac{f_{RR}}{1+f_{R}}}.
\end{equation}
Notice that for the $f(R)$ model, $G_{eff}$ depends on not only the scale factor but also the comoving wavenumber $k$.
As mentioned in \cite{fRkvalue}, the subhorizon approximation cannot be satisfied and
the non-linear effects are obvious in scale smaller than $k=0.2h{\rm Mpc^{-1}}$,
while the current observations are not so accurate for scale larger than $k=0.01h{\rm Mpc^{-1}}$.
Therefore, for simplicity, we just take $k=0.1h{\rm Mpc^{-1}}$.

For the $f(R)$ models, we consider two metric forms,
proposed by Hu-Sawicki \cite{fR1} and Starobinsky \cite{fR2}, respectively.
We will call them $f(R)_{HS}$ and $ f(R)_{St}$ in the following context.
Both of them can satisfy the cosmological and local gravity constraints \cite{Tsujikawa1_fR}.
In the following two subsections,
we will discuss in detail the explicit formulas and cosmological interpretations of these two models.

\subsubsection{The $f(R)_{HS}$ model }

The model proposed by Hu-Sawicki \cite{fR1} has the following form of $f(R)$,
\begin{equation} \label{eq:f(R)_HS}
f(R) = - \mu R_{c}\frac{(R/R_{c})^{2n}}{(R/R_{c})^{2n}+1},
\end{equation}
where $\mu$ and $n$ are positive numbers,
and $R_c$ is the order of the present Ricci scalar $R_0$.
In this paper, we take Hu-Sawicki's suggestion of $R_c = \tilde{\Omega}_{m0} H_0^2$,
and further set $\tilde{\Omega}_{m0} = 0.25$.
As shown in \cite{CapTsu08},
by using the constraints from the violations of weak and strong equivalence principles,
Capozziello and Tsujikawa give a bound $n>0.9$.
In practice \cite{fRkvalue}, $n$ is often treated as an integer.
For simplicity, we will focus on the case of  $n=1$.
The effects of different $n$ will be shown in Appendix B.
So actually for the $f(R)_{HS}$ model, we only have two free model parameters, i.e., $\mu$ and $\Omega_{m0}$.
Correspondingly, the Hubble parameter $H(z)$ satisfies the following equation:
\begin{eqnarray}
H^{2}
&&- \frac{1}{6} \left[ \mu R_{c}\frac{(R/R_{c})^{2n}}{(R/R_{c})^{2n}+1} \right] + \left\lbrace  \frac{ 2 \mu  n \left(\frac{R}{R_{c}}\right)^{2 n - 1}}{
   \left[\left(\frac{R}{R_{c}}\right)^{2 n}+1\right]^2} \right\rbrace
    (\frac{R}{6}-H^{2})  \nonumber \\
&&+ \left\lbrace  \frac{2 \mu  n R_{c}  \left[ \left(\frac{R}{R_{c}}\right)^{2 n}+2 n
   \left(\left(\frac{R}{R_{c}}\right)^{2 n}-1\right)+1 \right]
   \left(\frac{R}{R_{c}}\right)^{2 n}}{R^2
   \left[ \left(\frac{R}{R_{c}}\right)^{2 n}+1 \right]^3} \right\rbrace
R^{'}H^{2}  \nonumber \\
&&= \frac{8 \pi G}{3}(\rho_m + \rho_r ).
\end{eqnarray}
One can solve this equation numerically to obtain the evolution of $H(z)$.
From Eqs.(\ref{eq:fRGeff}) and (\ref{eq:f(R)_HS}),
the $G_{eff}$ of this $f(R)$ model can also be obtained.

\begin{figure}
\includegraphics[scale=0.58, angle=0]{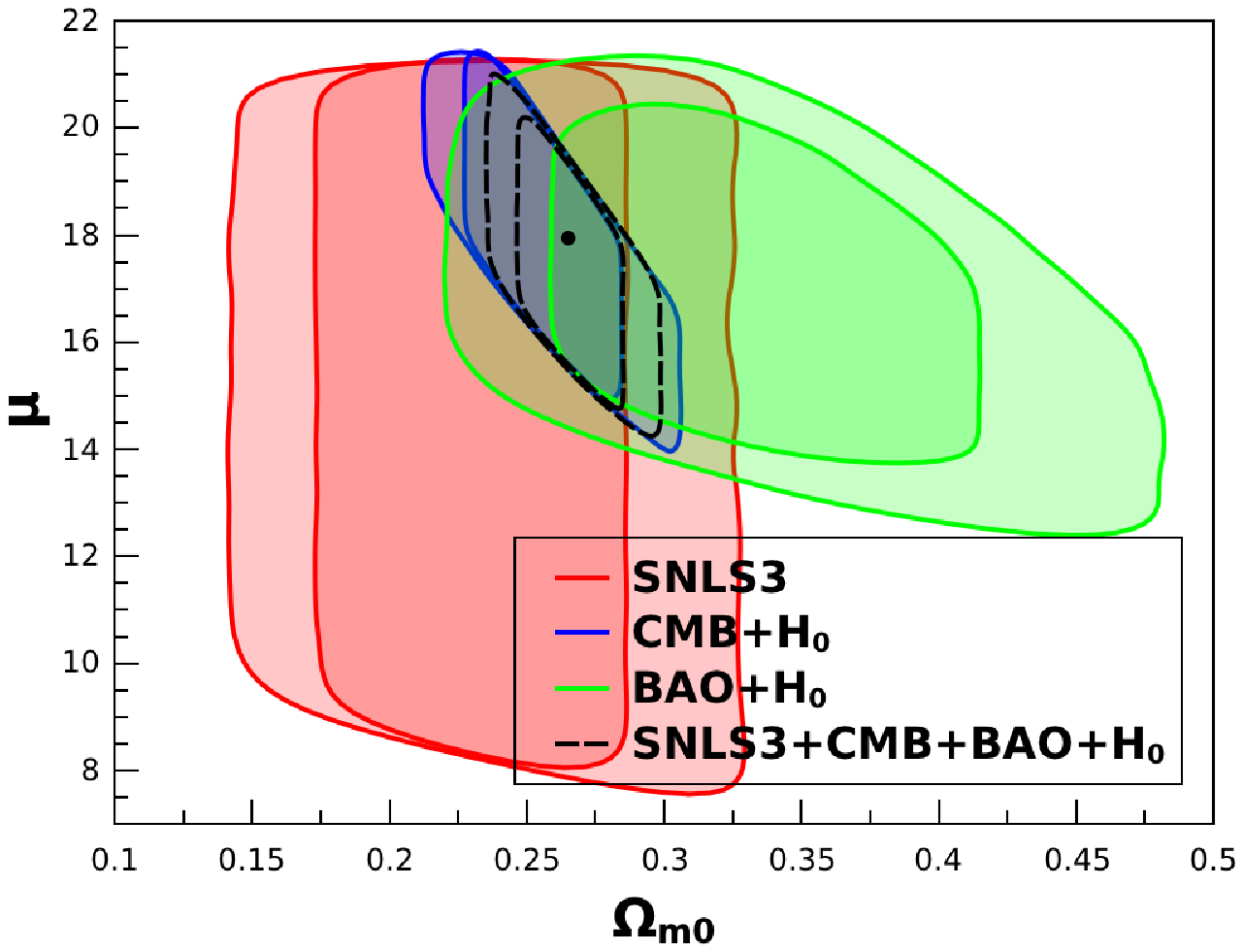}
\includegraphics[scale=0.58, angle=0]{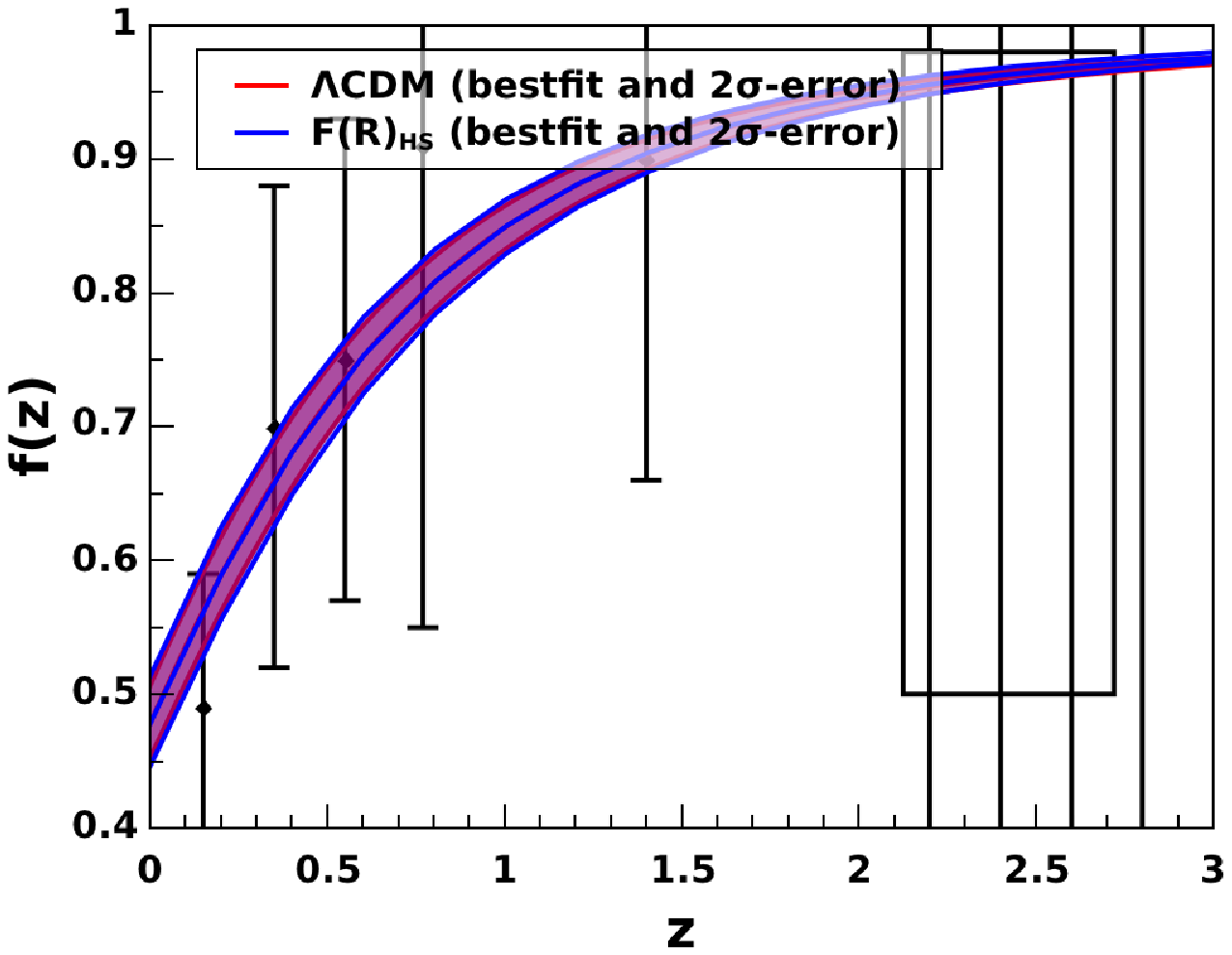}
\caption{\label{fRhsfg}
$Left\ Panel$: Probability contours at the $68.3\%$ and $95.4\%$ confidence levels
in $\Omega_{m0} \textendash \mu$ plane, for the $f(R)_{HS}$ model.
$Right\ Panel$: the evolutions of $f(z)$ predicted by the $\Lambda$CDM and $f(R)_{HS}$ models.
The best-fit values and $2\sigma$ errors regions determined by SNIa+CMB+BAO+$H_{0}$ analysis are shown.
In the plot of the $f(R)$ model, $k=0.1h{\rm Mpc^{-1}}$ is chosen to satisfy the subhorizon approximation.}
\end{figure}

Now let us discuss the cosmological constraints of the $f(R)_{HS}$ model.
In the left panel of Fig.\ref{fRhsfg}, we plot the contours of $68.3\%$ and $95.4\%$ confidence levels
in the $\Omega_{m0} \textendash \mu$ plane for the $f(R)_{HS}$ model in the case of $n=1$.
Constraints from SNLS3, CMB+$H_0$, BAO+$H_0$, and SNLS3+CMB+BAO+$H_0$ are shown in contours with different colors.
At the $68.3\%$ and $95.4\%$ confidence levels, we obtain
\begin{equation}
\Omega_{m0} = 0.2648^{+0.0212}_{-0.0192}\ ^{+0.0359}_{-0.0305},
\ \ \ \mu = 17.2603^{+2.3439}_{-3.1935} \ ^{+3.0800}_{-3.7323},
\ \ \ h = 0.7120^{+0.0164}_{-0.0165}\ ^{+0.0267}_{-0.0270}.
\end{equation}

In the right panel of Fig.\ref{fRhsfg}, we plot the evolutions of $f(z)$
predicted by the $\Lambda$CDM model and the $f(R)_{HS}$ model
along with the observed data of growth factor $f_{obs}$.
These two models have similar evolutions of $f(z)$,
and it is quite difficult for us to distinguish these two models from the current growth factor data.

\subsubsection{The $f(R)_{St}$ model }

Starobinsky also proposed a famous viable $f(R)$ model \cite{fR2}, in which
\begin{equation}\label{eq:f(R)_St}
f(R)= - \lambda R_{s}[1-(1+\frac{R^2}{R_{s}^2})^{-n}],
\end{equation}
where $\lambda$ and $n$ are positive numbers.
The same as the $f(R)_{HS}$ model, we choose $R_s = 0.25 H_0^2$.
As shown in \cite{CapTsu08},
this model also has a bound $n>0.9$.
So in this work, this model is also treated as a two-parameter model with parameter $\lambda$ and $\Omega_{m0}$.
Combining the above equation with Eqs. (\ref{eq:Friedmann1fR}-\ref{eq:fRGeff}),
the evolution of Hubble parameter $H(z)$ can be obtained by numerically solving the following equation:
\begin{eqnarray}
H^{2}
&&- \frac{1}{6} \left[ \lambda R_{s}\frac{(R/R_{s})^{2n}}{(R/R_{s})^{2n}+1} \right]
   +  \left[ \frac{2 \lambda  n R
   \left(\frac{R^2}{R_{s}^2}+1\right)^{-n-1}}{R_{s}} \right]
    (\frac{R}{6}-H^{2} ) \nonumber \\
&&+ \left\lbrace  \frac{2 \lambda n R_{s} \left(\frac{R^2}{R_{s}^2}+1\right)^{-n}
   \left[ (2 n+1) R^2 - R_{s}^2 \right]  }{\left(R^2+R_{s}^2\right)^2} \right\rbrace
R^{'}H^{2} \nonumber \\
&&= \frac{8 \pi G}{3}(\rho_m + \rho_r ).
\end{eqnarray}
Substituting Eq.(\ref{eq:f(R)_St}) into Eq.(\ref{eq:fRGeff}),
one can also obtain the $G_{eff}$ of this $f(R)$ model.

\begin{figure}
\includegraphics[scale=0.59, angle=0]{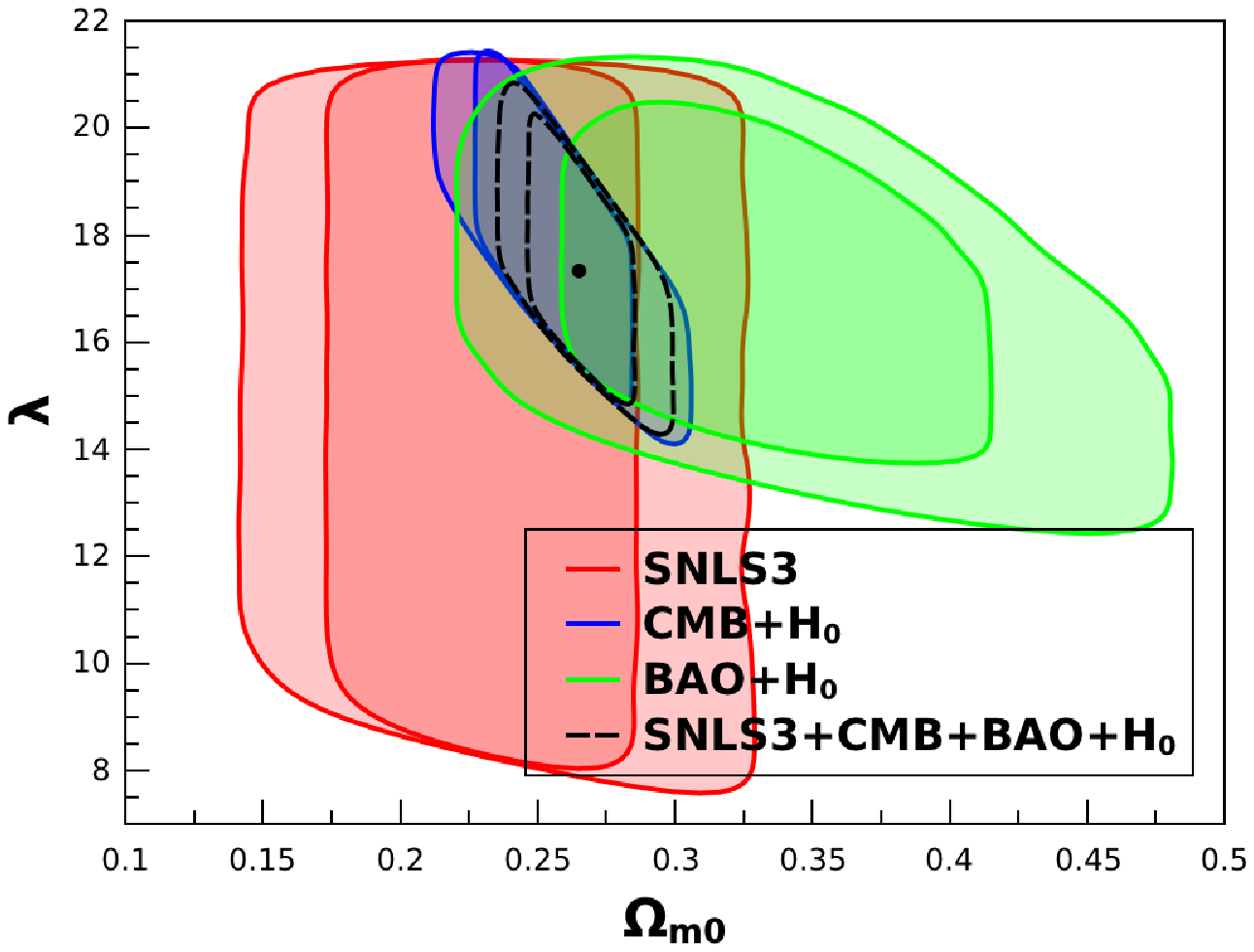}
\includegraphics[scale=0.58, angle=0]{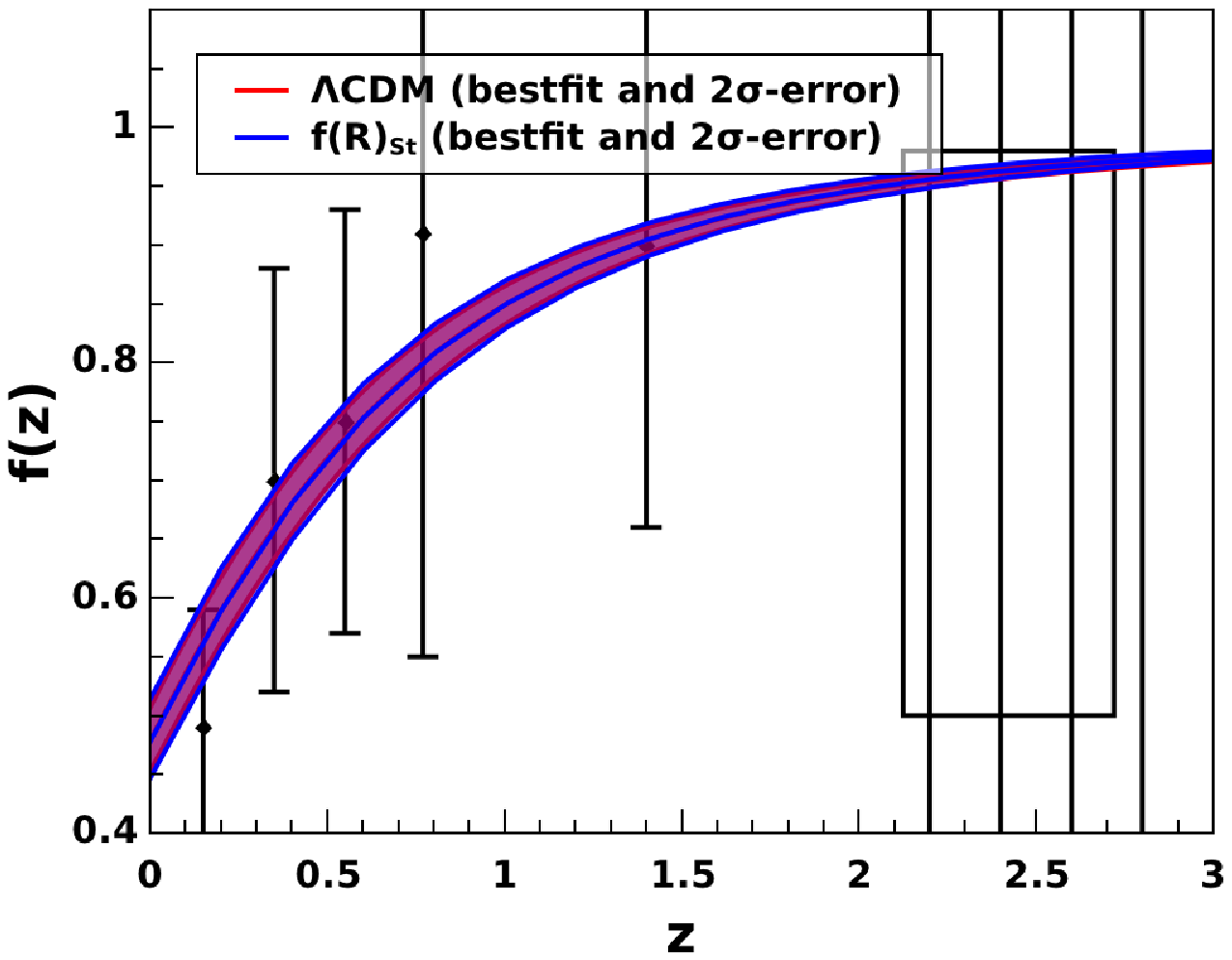}
\caption{\label{fRstfg}
$Left\ Panel$: Probability contours at the $68.3\%$ and $95.4\%$ confidence levels
in $\Omega_{m0} \textendash \lambda$ plane, for the $f(R)_{St}$ model.
$Right\ Panel$: The evolutions of $f(z)$ predicted by the $\Lambda$CDM and $f(R)_{St}$ models.
The best-fit values and $2\sigma$ errors regions determined by SNIa+CMB+BAO+$H_{0}$ analysis are shown.
Again, we choose $k=0.1h{\rm Mpc^{-1}}$ in order to satisfy the subhorizon approximation.}
\end{figure}

In the left panel of Fig.\ref{fRstfg}, we plot the contours of $68.3\%$ and $95.4\%$ confidence levels
in the $\Omega_{m0} \textendash \lambda$ plane for the $f(R)_{St}$ model with $n=1$.
Constraints from SNLS3, CMB+$H_0$, BAO+$H_0$, and SNLS3+CMB+BAO+$H_0$ are shown in contours with different colors.
At the $68.3\%$ and $95.4\%$ confidence levels,
\begin{equation}
\Omega_{m0} = 0.2646^{+0.0215}_{-0.0190}\ ^{+0.0363}_{-0.0302},
\ \ \ \lambda = 17.2771^{+3.0432}_{-2.6320} \ ^{+3.6433}_{-3.2810},
\ \ \ h = 0.7122^{+0.0160}_{-0.0167}\ ^{+0.0263}_{-0.0271}.
\end{equation}

In the right panel of Fig.\ref{fRstfg}, we plot the evolutions of $f(z)$
predicted by th $\Lambda$CDM model and the $f(R)_{St}$ model
along with the observational data of growth factor $f_{obs}$.
Again, we obtain similar results:
the evolution of the growth factor in the $f(R)_{St}$ model is
indistinguishable from that of the $\Lambda$CDM model.

\subsection{Summary}\label{Summary}

A brief summary of our results is shown in Table \ref{SumofResults},
where the models, model parameters, number of free model parameters $n_{p}$, $\chi^2_{min}$\ s,
$\Delta$AIC and $\Delta$BIC are given.
The $\Delta$AIC and $\Delta$BIC are defined as,
\begin{eqnarray}
\Delta {\rm AIC} &=& {\rm AIC_{model}-AIC_{\Lambda CDM}},\\
\Delta {\rm BIC} &=& {\rm BIC_{model}-BIC_{\Lambda CDM}}.
\end{eqnarray}
The nuisance parameters $h$, $\alpha$ and $\beta$
are actually not model parameters with significant meanings,
and hence are not listed in the table.

\begin{table}
\begin{center}
\caption{\label{SumofResults} The Summary of Results for five MG models and the $\Lambda$CDM model.}
\begin{tabular}{|c|c|c|c|c|c|}
\hline
Models & Model parameters & $n_{p}$ & $\chi_{min}^{2}$ & $\Delta$AIC & $\Delta$BIC \tabularnewline
\hline
$\Lambda$CDM & $\Omega_{m0}$ & $1$  & $424.911$ & $0$ & $0$ \tabularnewline
\hline
DGP & $\Omega_{m0}$ & $1$  & $470.231$  & $45.32 $ & $45.32$ \tabularnewline
\hline
$f(T)_{PL}$ & $n,\ \Omega_{m0}$ & $2$ & $423.506$ & $0.595 $ & $4.765$ \tabularnewline
\hline
$f(T)_{EXP}$ & $p,\ \Omega_{m0}$ & $2$  & $423.275$ & $0.364 $ & $4.534$ \tabularnewline
\hline
$f(R)_{HS}$ & $\mu,\ \Omega_{m0}$ & $2$  & $424.901$ & $1.990 $ & $5.434$ \tabularnewline
\hline
$f(R)_{St}$ & $\lambda,\ \Omega_{m0}$ & $2$ & $424.903$ & $1.992$ & $6.161$ \tabularnewline
\hline
\end{tabular}
\end{center}
\begin{footnotesize}
Note: The nuisance parameters $h$, $\alpha$, $\beta$
used in the analysis are actually not model parameters with significant meanings,
so we do not list them in this table.
\end{footnotesize}
\end{table}

To make a comparison,
we also list the case of the $\Lambda$CDM model.
As shown in the Table \ref{SumofResults},
the values of $\Delta$AIC and the $\Delta$BIC of DGP model are both quite lager than 6.
This means that the DGP model is strongly disfavored by the data.
Other MG models do not yield to remarkable reductions of the $\chi^2_{min}$,
and give slightly larger AIC and BIC values compared with the $\Lambda$CDM model.
This indicates that the $\Lambda$CDM model is still more favored by the current data.
This result is consistent with some previous works
\cite{AIC,BIC,cosmologyICGodlowski,cosmologyICBiesiada,cosmologyICMagueijo,DEManyModelsDRubin,ModelComp}.

\section{Concluding Remarks}\label{Discussion And Conclution}

In this work, we test 5 MG models by using the current cosmological observations.
Utilizing the observational data of the cosmic expansion history,
including the recently released SNLS3 type Ia supernovae sample,
the cosmic microwave background anisotropy data from the WMAP7 observations,
the baryon acoustic oscillation results from the SDSS DR7
and the latest Hubble constant measurement utilizing the WFC3 on the HST,
we constrain the parameter spaces of these models.
Then, by plotting the evolutions of these models' growth factor,
we further compare the theoretical predictions of these MG models
with the current growth factor data.
It is found that these MG models do not lead to appreciable reductions of the $\chi^2_{min}$,
and give larger AIC and BIC values compared with the $\Lambda$CDM model.
In addition, based on the current growth factor data,
these MG models are difficult to be distinguished from the $\Lambda$CDM model,
so further growth factor data is needed.

\section*{Acknowledgements}
This work was supported by the NSFC grant No.10535060/A050207,
a NSFC group grant No.10821504.
QGH was also  supported by the project of Knowledge Innovation
Program of Chinese Academy of Science and a grant from NSFC (Grant No. 10975167).

\bigskip

\section*{Appendix A. Initial conditions of the matter density perturbation equation}

In this Appendix, we explain the reason of taking initial conditions $f(z=30)=1$
when solving the Eq(\ref{eq:deltagf}) numerically.

Starting from Eq.(\ref{eq:deltarho}) and changing variables from $t$ to $a$, one can obtain
\begin{equation}\label{eq:delt-rho-a}
\frac{d^2\delta}{da^2}+(\frac{3}{a}+\frac{d\ln H}{da})\frac{d\delta}{da}
-\frac{4\pi G_{eff} \rho_{m}}{a^2H^2}\delta=0.
\end{equation}
To solve this equation numerically,
we take the initial condition at high-$z$ era, e.g. $z=30$ here,
since $G_{eff} = G$ is satisfied precisely at high-$z$ era for all models considered in this paper.

In addition, in the high-$z$ regime, the universe is at the matter-dominated stage,
thus we have the Friedmann equation:
\begin{equation}
H^2=\frac{8\pi G}{3}\rho_{m}.
\end{equation}

Then, Eq.(\ref{eq:delt-rho-a}) becomes
\begin{equation}\label{eq:delt-rho-a2}
\frac{d^2\delta}{da^2}+\frac{3}{2a}\frac{d\delta}{da}-\frac{3}{2a^2}\delta=0.
\end{equation}

Assuming the solution of the above equation takes the form
$\delta=\delta_{const}a^m$, and then substituting this form into Eq.(\ref{eq:delt-rho-a2}),
one finally gets the function's general solution
\begin{equation}
\delta=C_1 a+C_2 a^{-3/2}.
\end{equation}

Since $\delta$ is quite small in high-$z$ regime, the acceptable solution is
\begin{equation}
\delta=C_1 a.
\end{equation}

Thus, we finally obtain the initial condition of Eq.(\ref{eq:deltagf})
\begin{equation}
f = \frac{d\ln \delta}{d\ln a}  = 1
\end{equation}
at redshift $z=30$.

\section*{Appendix B. The effects of $n$ in the two $f(R)$ models }

In this Appendix, we show that
the effects of $n$ on the cosmological interpretations of the $f(R)$ models are rather small.

Let us consider the $f(R)_{HS}$ model.
The best-fit values of the cosmological parameters $\Omega_{m0}$ and $h$,
their $68.3\%$ and $95.4\%$ confidence region,
and the corresponding $\chi^{2}_{min}$ can be obtained from the joint analysis of the SNLS3+CMB+BAO+$H_0$ data.
When we take $n=1$, we find the following results,
\begin{eqnarray}
\Omega_{m0} = 0.2648^{+0.0212}_{-0.0192}\ ^{+0.0359}_{-0.0305},
\ \ h = 0.7120^{+0.0164}_{-0.0165}\ ^{+0.0267}_{-0.0270}, 
\ \ \chi^{2}_{min} &=& 424.901,
\end{eqnarray}
while $n=2$ yields to
\begin{eqnarray}
\Omega_{m0} = 0.2648^{+0.0214}_{-0.0190}\ ^{+0.0357}_{-0.0305} ,
\ \ h = 0.7121^{+0.0160}_{-0.0165}\ ^{+0.0265}_{-0.0268}, 
\ \ \chi^{2}_{min} &=& 424.913.
\end{eqnarray}
Clearly, the above two sets of results are very close to each other.

\begin{figure}
\includegraphics[scale=0.58, angle=0]{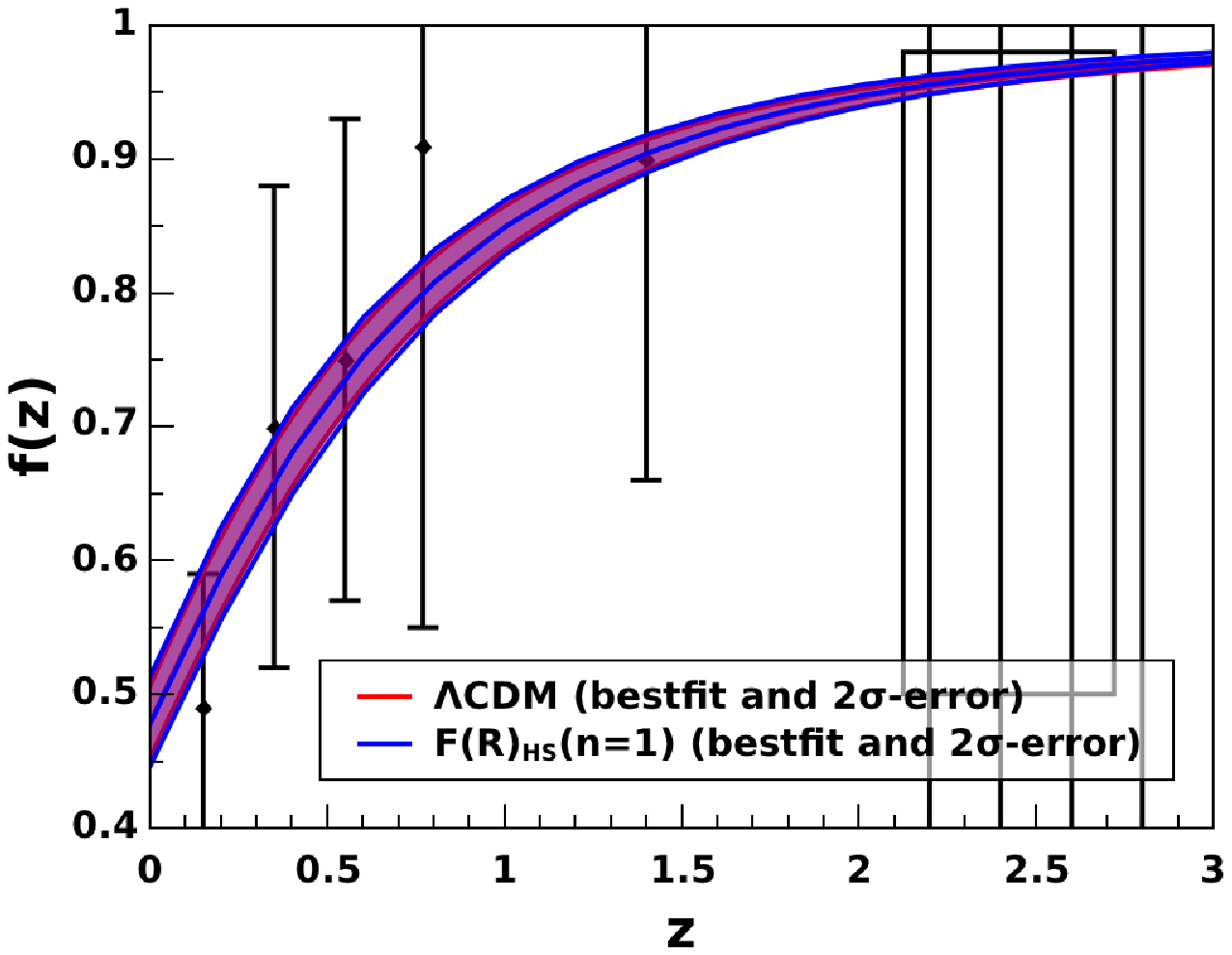}
\includegraphics[scale=0.58, angle=0]{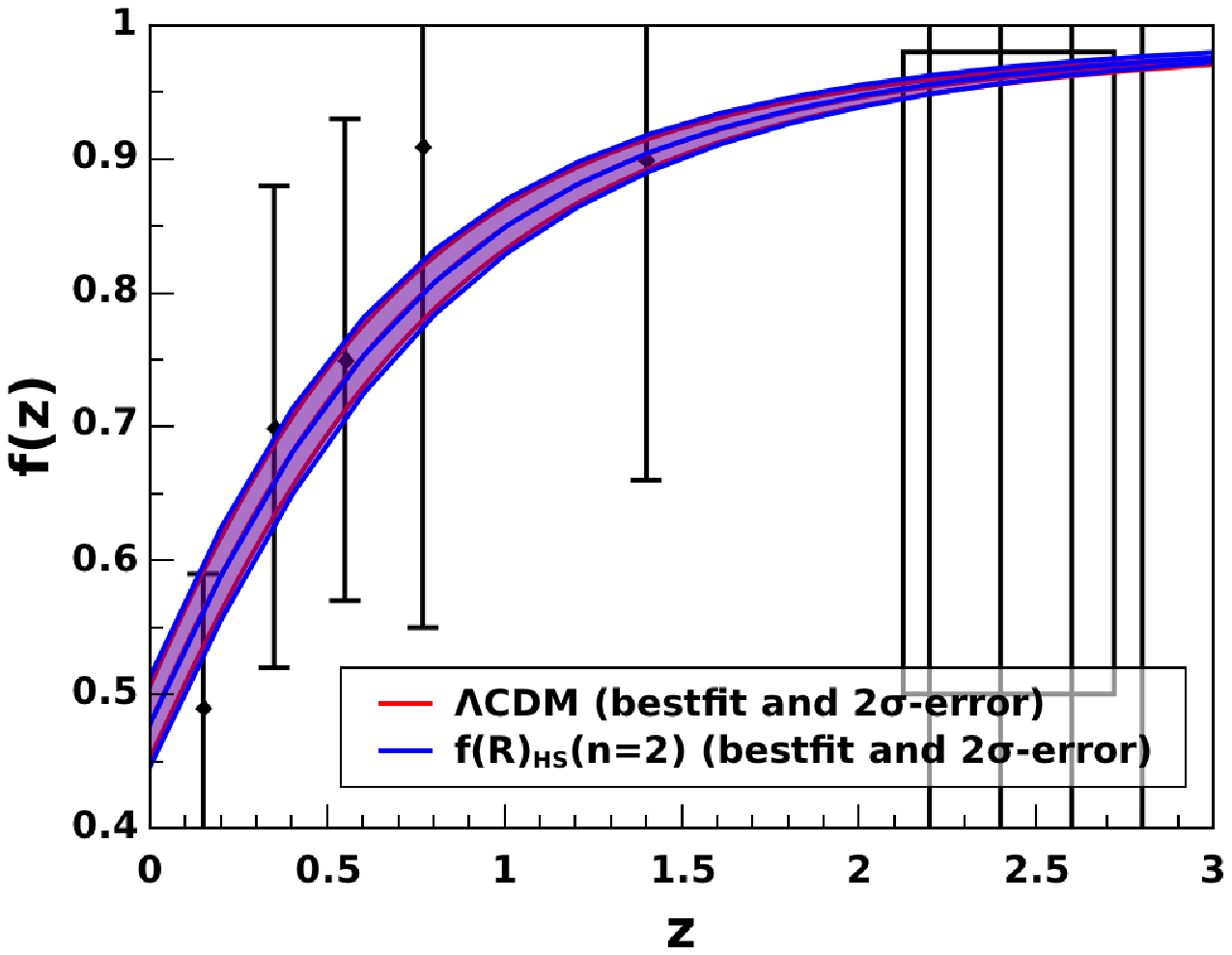}
\caption{\label{fRhsn12}
The evolutions of $f(z)$ predicted by the $\Lambda$CDM and $f(R)_{HS}$ models
in the cases of $n=1$ and $n=2$ respectively.
The best-fit values and $2\sigma$ errors regions determined by SNIa+CMB+BAO+$H_{0}$ analysis are shown.
Note that, $k=0.1h{\rm Mpc^{-1}}$ was chosen in order to satisfy the subhorizon approximation.}
\end{figure}

In Fig.\ref{fRhsn12}, we also plot the evolutions of $f(z)$ predicted by
the $\Lambda$CDM model and the $f(R)_{HS}$ model with $n=1$ and $n=2$,
along with the observed data of growth factor $f_{obs}$.
Also, it can be seen that the results of $n=1$ and $n=2$ are similar to each other.

In all, it is clear that different $n$ give quite similar results
of the constraints on $\Omega_{m0}$ and $h$, the $\chi_{min}$s, and the evolutions of $f(z)$.
That is, the effects of $n$ on the cosmological interpretations of the $f(R)_{HS}$ model are rather small.
In the $f(R)_{St}$ model, the result is similar.
Therefore, as mentioned above, it is unnecessary for us to treat $n$
as a free model parameter for the two $f(R)$ models.

\end{document}